 \documentclass[11pt]{article}
\usepackage[english]{babel}
\usepackage{graphicx}
\usepackage{color}

\usepackage{epsfig}
   \usepackage{graphics,color,pstricks,pst-plot,pst-node,pst-coil}
   \usepackage[cp1251]{inputenc}

\begin{document}

\title{{\bf Elementary  particles 
  in the early Universe }
}

\author{{\bf  N.~A.~Gromov }\\
 Komi Science Center of the Ural Division\\
  of the Russian Academy of Sciences,
 Syktyvkar, 167982, Russia.\\
Electronic address: gromov@dm.komisc.ru}
 
\date{}

\maketitle

\begin{abstract}

The high-temperature limit of   Standard Model  generated by the contractions of gauge groups is discussed.
Contraction parameters of gauge group $SU(2)$ of  Electroweak Model and gauge group $SU(3)$ of  Quantum Chromodynamics are taken identical and tending to zero when temperature  increase.
Properties of the elementary particles change  drastically
at the infinite temperature limit: all particles lose masses, all quarks are monochromatic. Electroweak interactions become long-range
and are mediated by the neutral currents. 
Particles of different kind do not interact.
It looks like some stratification  with only one sort of particles in each stratum.
The Standard Model passes in this limit through several stages, which are distinguished by the powers of contraction parameter.
For any stage the intermediate models  are constructed and the exact expressions for the respective Lagrangians are presented.
The developed approach describes the evolution of  Standard Model in the early Universe from the Big Bang up to the end of several nanoseconds.
\end{abstract}



\section{Introduction}



Modern theory of elementary particles known as  Standard Model (SM) consist of  Electroweak Model (EWM), which unified electromagnetic and weak interactions, as well as  Quantum  Chromodynamics (QCD), describing their strong interactions.
Standard Model gives a good description of the experimental data and was recently  confirmed by discovering of Higgs boson at LHC. Standard Model is a gauge theory with
$SU(3)\times SU(2)\times U(1)$
gauge group, which is direct product of a simple groups.
 The operation of group contraction (or limit transition) \cite{IW-53} well known in physics transforms a simple group to a non-semisimple  one.
For a symmetric physical system the contraction of its symmetry group means a transition to some limit state. In the case of a complicated physical system the investigation of its limit states under the limit values of some of its parameters  enables to better understand the system behavior. 

For the modified Electroweak Model with the contracted gauge group $ SU(2;j)\times U(1)$ 
it was demonstrated \cite{Gr-12,Gr-2012,Gr-2013},
that the contraction parameter is connected with system  energy, so its zero limit corresponds to the low-energy limit of  Electroweak Model.
The alternative rescaling of the gauge group and the field space gives  the  high-temperature   limit of Electroweak Model.

In the broad sense of the word  deformation is an operation inverse to contraction. The non-trivial deformation of some algebraic structure generally means 
its non-evident generalization.
Quantum groups \cite{FRT}, which are simultaneously non-commutative and non-cocommutative Hopf algebras, present a good example of similar generalization since previously
Hopf algebras with only one of these properties were known.
But when  the contraction of  some mathematical or physical structure is performed one can reconstruct  the initial structure by the deformation in the narrow sense moving back along the contraction way.

We use this method in order to re-establish the evolution of elementary particles and their interactions in the early Universe. We are   based on the modern knowledge of the particle world which is concentrated in  Standard Model.
In this paper  we investigate the high-temperature limit of  Standard Model generated by contraction of the gauge groups $SU(2)$ and $SU(3)$.
Similar very high temperatures  can exist in the early Universe
after inflation and reheating on the first stages of the Hot Big Bang \cite{GR-2011}.
At these times  the elementary particles demonstrate rather unusual properties. 
 It appears that the SM Lagrangian falls to  a number of terms which are distinguished by the powers of the contraction parameter $ \epsilon \rightarrow 0 $.
As far as the  temperature in the hot Universe is connected with its age, then moving forward in time, i.e. back to high-temperature  contraction, we conclude that after the Universe creation elementary particles and their interactions pass a number of stages in their evolution from infinite temperature state up to SM state.
These stages   of quark-gluon plasma formation and   color symmetries restoration  are distinguished by the powers of contraction parameter and consequently by times of its creations.
From the contraction of  Standard Model we can classify the stages in time as earlier-later, but we can not
determine their absolute date. To estimate the absolute date we use  additional assumptions.

\section{ High-Temperature Lagrangian of EWM} 

It was shown in papers \cite{Gr-12,Gr-2012,Gr-2013},
 that in zero energy limit
the gauge  group $SU(2;j)$ 
and the fundamental representation space ${\bf C}_2(j)$   are transformed as follows
$$
z'(j)=
\left(\begin{array}{c}
jz'_1 \\
z'_2
\end{array} \right)
=\left(\begin{array}{cc}
	\alpha & j\beta   \\
-j\bar{\beta}	 & \bar{\alpha}
\end{array} \right)
\left(\begin{array}{c}
jz_1 \\
z_2
\end{array} \right)
=u(j)z(j), \quad
z(j)= \left(
\begin{array}{c}
	j\nu_l \\
	e_l
\end{array} \right),\ldots,
\left(
\begin{array}{c}
	ju_l\\
	d_l
\end{array} \right),\ldots,
$$
\begin{equation}
\det u(j)=|\alpha|^2+j^2|\beta|^2=1, \quad u(j)u^{\dagger}(j)=1,
\label{eq5-21}
\end{equation} 
 so for $j \rightarrow 0$
the first components of the lepton and quark doublets become infinitely small in comparison to
their  second components. On the contrary, when energy (temperature)  increases the first components of the  doublets become
greater than their  second ones.  In the infinite temperature  limit the second components of the lepton and quark doublets will be infinitely small as compared to their  first components. To describe this limit we introduce \cite{Gr-2014,Gr-2015} {\it new  contraction parameter} $\epsilon$ and {\it new    consistent rescaling} of the group $SU(2;\epsilon)$  and the  space ${\bf C}_2(\epsilon)$
in the form 
 $$
z'(\epsilon)=
\left(\begin{array}{c}
 z'_1 \\
\epsilon z'_2
\end{array} \right)
=\left(\begin{array}{cc}
	\alpha & \epsilon\beta   \\
-\epsilon\bar{\beta}	 & \bar{\alpha}
\end{array} \right)
\left(\begin{array}{c}
z_1 \\
\epsilon z_2
\end{array} \right)
=u(\epsilon)z(\epsilon),  
$$
\begin{equation}
\det u(\epsilon)=|\alpha|^2+\epsilon^2|\beta|^2=1, \quad u(\epsilon)u^{\dagger}(\epsilon)=1.
\label{5}
\end{equation}
Hermite form
\begin{equation}
z^\dagger(\epsilon)z(\epsilon)=|z_1|^2+\epsilon^2|z_2|^2
\label{eq5-31}
\end{equation}
remain invariant under contraction $\epsilon \rightarrow 0$.

In the  contraction scheme (\ref{5})  the standard boson fields and left lepton and quark fields are transformed as follows 
$$
W_{\mu}^{\pm} \rightarrow \epsilon W_{\mu}^{\pm}, \;\; Z_{\mu} \rightarrow Z_{\mu},\; \;A_{\mu} \rightarrow A_{\mu}.
$$
\begin{equation}
 	 e_{l} \rightarrow \epsilon e_{l},  \quad  d_{l} \rightarrow \epsilon d_{l}, \quad
 	 \nu_l \rightarrow \nu_l, \quad  	u_l \rightarrow u_l.
\label{6}
\end{equation}
The next reason for inequality of the first and second doublet components is the special mechanism of spontaneous symmetry breaking, which is used to generate  mass of vector bosons and other elementary particles
of the model.
In this mechanism one of Lagrangian 
ground states
$
  \phi^{vac}=\left(\begin{array}{c}
	0  \\
	v
\end{array} \right) \;
$
is taken as vacuum of the model and then small field excitations $ v+\chi(x) $ with respect to this vacuum are regarded.
So Higgs boson field $ \chi $ and constant $ v $ are multiplied by $\epsilon$.
As far as masses of all particles are proportionate to $ v $ we obtain the following transformation rule 
\begin{equation}
 	\chi  \rightarrow \epsilon \chi,  \quad  v \rightarrow \epsilon v, \quad 
 m_p \rightarrow \epsilon m_p, \quad p={\chi}, W, Z, e, u, d.
\label{7}
\end{equation}

After  transformations (\ref{6}), (\ref{7})
the EWM boson Lagrangian \cite{R-99,PS-95} can be represented in the form
\begin{equation}
 L_B(\epsilon)= - \frac{1}{4}{\cal Z}_{\mu\nu}^2 - \frac{1}{4}{\cal F}_{\mu\nu}^2 + \epsilon^2 L_{B,2}
 + \epsilon^3 gW_\mu^{+}W_\mu^{-}\chi  + \epsilon^4 L_{B,4},
\label{8}
\end{equation}
where
\begin{equation} 
 L_{B,4}= m_W^2W_\mu^{+}W_\mu^{-} -\frac{1}{2}m_{\chi}^2\chi^2 -\lambda v \chi^3  - \frac{\lambda}{4} \chi^4 
 +\frac{g^2}{4}\left(W_\mu^{+}W_\nu^{-} - W_\mu^{-}W_\nu^{+}\right)^2 +
\frac{g^2}{4}W_\mu^{+}W_\nu^{-}\chi^2,
\label{8-1}
\end{equation}
$$
L_{B,2}= \frac{1}{2}\left(\partial_\mu\chi \right)^2
 + \frac{1}{2}m_Z^2\left(Z_\mu\right)^2 -\frac{1}{2}{\cal W}_{\mu\nu}^{+}{\cal W}_{\mu\nu}^{-} 
 +\frac{g m_z}{2\cos \theta_W} \left(Z_{\mu}\right)^2 \chi +
 \frac{g^2 }{8\cos^2\theta_W} \left(Z_{\mu}\right)^2 \chi^2 -
 $$
 $$
-2ig\left(W_\mu^{+}W_\nu^{-} - W_\mu^{-}W_\nu^{+}\right)
\Bigl( {\cal F}_{\mu\nu}\sin \theta_W + {\cal Z}_{\mu\nu}\cos \theta_W \Bigr)  -
$$
$$
-\frac{i}{2}e \left[A_{\mu}\left({\cal W}_{\mu\nu}^{+}W_\nu^{-} - {\cal W}_{\mu\nu}^{-}W_\nu^{+}\right) +
 \frac{i}{2}e A_{\nu}\left({\cal W}_{\mu\nu}^{+}W_\mu^{-} - {\cal W}_{\mu\nu}^{-}W_\mu^{+}\right) \right] -
$$
$$
-\frac{i}{2}g\cos \theta_W  \left[Z_{\mu}\left({\cal W}_{\mu\nu}^{+}W_\nu^{-} - {\cal W}_{\mu\nu}^{-}W_\nu^{+}\right) - \right.
$$
$$
\left.  -Z_{\nu}\left({\cal W}_{\mu\nu}^{+}W_\mu^{-} - {\cal W}_{\mu\nu}^{-}W_\mu^{+}\right) \right] 
-\frac{e^2}{4} \left\{
\left[\left(W_{\mu}^{+}\right)^2 + \left(W_{\mu}^{-}\right)^2\right](A_{\nu})^2- \right.
$$
$$
-2\left(W_\mu^{+}W_\nu^{+} + W_\mu^{-}W_\nu^{-} \right)A_{\mu}A_{\nu} + 
 \left.
\left[\left(W_{\nu}^{+}\right)^2 + \left(W_{\nu}^{-}\right)^2\right](A_{\mu})^2
\right\} -
$$
$$
-\frac{g^2}{4}\cos\theta_W \left\{
\left[\left(W_{\mu}^{+}\right)^2 + \left(W_{\mu}^{-}\right)^2\right](Z_{\nu})^2 - \right.
$$
$$
\left. 
-2\left(W_\mu^{+}W_\nu^{+} + W_\mu^{-}W_\nu^{-} \right)Z_{\mu}Z_{\nu} +
\left[\left(W_{\nu}^{+}\right)^2 + \left(W_{\nu}^{-}\right)^2\right](Z_{\mu})^2
\right\} -
$$
\begin{equation} 
 -eg\cos\theta_W \biggl[
W_\mu^{+}W_\mu^{-}A_{\nu}Z_{\nu} +
W_\nu^{+}W_\nu^{-}A_{\mu}Z_{\mu} 
-\frac{1}{2}\left(W_\mu^{+}W_\nu^{-} + W_\nu^{+}W_\mu^{-} \right)\left(A_{\mu}Z_{\nu} + A_{\nu}Z_{\mu}\right)
\biggr].
\label{8-2}
\end{equation}

The lepton  Lagrangian  in terms of electron and neutrino fields takes the form
$$
L_{L}(\epsilon)= L_{L,0} + \epsilon^2 L_{L,2} =
$$
$$
=\nu_l^{\dagger}i\tilde{\tau}_{\mu}\partial_{\mu}\nu_l +
e_r^{\dagger}i\tau_{\mu}\partial_{\mu}e_r  +g'\sin \theta_w e_r^{\dagger}\tau_{\mu}Z_{\mu}e_r 
- g'\cos \theta_w e_r^{\dagger}\tau_{\mu}A_{\mu}e_r
 + \frac{g}{2\cos \theta_w} \nu_l^{\dagger}\tilde{\tau}_{\mu}Z_{\mu}\nu_l +
 $$
 $$
 +\epsilon^2\biggl\{ e_l^{\dagger}i\tilde{\tau}_{\mu}\partial_{\mu}e_l
-m_e(e_r^{\dagger}e_l + e_l^{\dagger} e_r)+
 \frac{g\cos 2\theta_w}{2\cos \theta_w}e_l^{\dagger}\tilde{\tau}_{\mu}Z_{\mu}e_l -
$$
\begin{equation}%
-ee_l^{\dagger}\tilde{\tau}_{\mu}A_{\mu}e_l 
+\frac{g}{\sqrt{2}}\left( \nu_l^{\dagger}\tilde{\tau}_{\mu}W_{\mu}^{+}e_l +
 e_l^{\dagger}\tilde{\tau}_{\mu}W_{\mu}^{-}\nu_l\right)\biggr\}.
\label{9}
\end{equation}

The quark  Lagrangian  in terms of u- and d-quarks  fields
can be written as
\begin{equation}
L_{Q}(\epsilon)= L_{Q,0} - \epsilon \, m_u(u_r^{\dagger}u_{l} + u_{l}^{\dagger}u_r) + \epsilon^2 L_{Q,2},
\label{12-1}
\end{equation}
where
$$
L_{Q,0}
=d_r^{\dagger}i\tau_{\mu}\partial_{\mu}d_r
+u_{l}^{\dagger}i\tilde{\tau}_{\mu}\partial_{\mu}u_{l}
+u_r^{\dagger}i\tau_{\mu}\partial_{\mu}u_r -
$$
$$
-\frac{1}{3}g'\cos\theta_w d_r^{\dagger}\tau_{\mu}A_{\mu}d_r
+ \frac{1}{3}g'\sin\theta_w d_r^{\dagger}\tau_{\mu}Z_{\mu}d_r
+\frac{2e}{3}u_{l}^{\dagger}\tilde{\tau}_{\mu}A_{\mu}u_{l} +
$$
\begin{equation} 
+\frac{g}{\cos \theta_w}\left(\frac{1}{2}-\frac{2}{3}\sin^2\theta_w\right) u_{l}^{\dagger}\tilde{\tau}_{\mu}Z_{\mu}u_{l} 
+\frac{2}{3}g'\cos\theta_w u_r^{\dagger}\tau_{\mu}A_{\mu}u_r
-\frac{2}{3}g'\sin\theta_w u_r^{\dagger}\tau_{\mu}Z_{\mu}u_r,
\label{12-2}
\end{equation}


$$
L_{Q,2}=d_{l}^{\dagger}i\tilde{\tau}_{\mu}\partial_{\mu}d_{l}
- m_d(d_r^{\dagger}d_{l} + d_{l}^{\dagger}d_r)
-\frac{e}{3}d_{l}^{\dagger}\tilde{\tau}_{\mu}A_{\mu}d_{l} -
$$
\begin{equation} 
-\frac{g}{\cos \theta_w}\left(\frac{1}{2}-
\frac{2}{3}\sin^2\theta_w\right) d_{l}^{\dagger}\tilde{\tau}_{\mu}Z_{\mu}d_{l}
+\frac{g}{\sqrt{2}}\left[ u_{l}^{\dagger}\tilde{\tau}_{\mu}W^{+}_{\mu}d_{l} +
d_{l}^{\dagger}\tilde{\tau}_{\mu}W^{-}_{\mu}u_{l}\right].
\label{12}
\end{equation}

The complete  Lagrangian of the modified model is given by the sum
$
L(\epsilon)=L_B(\epsilon) + L_L(\epsilon) + L_Q(\epsilon)
$
and can be written in the form
\begin{equation}
L(\epsilon)=L_{\infty} + \epsilon L_1 + \epsilon^2 L_2 + \epsilon^3 L_3 + \epsilon^4 L_4.
\label{15}
\end{equation}
The contraction parameter is monotonous function   $\epsilon(T) $ of the  temperature  with the property
$\epsilon(T) \rightarrow 0 $ for $T \rightarrow \infty $.
Very high energies can exist in the early Universe just after its creation.

It is well known that to gain a  better understanding of a physical system it is useful to investigate its properties  for limiting values of  physical parameters.
It follows from the decomposition (\ref{15}) that there are five stages in evolution   of the Electroweak Model after the creation of the Universe which are distinguished by the powers of the contraction parameter $ \epsilon $.
This offers an opportunity for construction of intermediate limit models. 
One can take the Lagrangian $L_{\infty}$ for the initial limit system, then add
$L_1$ 
and obtain the second limit model with the Lagrangian
${\cal L}_1=L_{\infty} +  L_1 $. After that one can add
$ L_2 $ and obtain the third limit model
${\cal L}_2=L_{\infty} + L_1 +  L_2 $.
The last  limit model has the Lagrangian
${\cal L}_3=L_{\infty} +  L_1 +  L_2 +  L_3 $.

 In the infinite temperature   limit ($\epsilon =0$) Lagrangian (\ref{15}) is equal to
$$
L_{\infty}= - \frac{1}{4}{\cal Z}_{\mu\nu}^2 - \frac{1}{4}{\cal F}_{\mu\nu}^2 +
\nu_l^{\dagger}i\tilde{\tau}_{\mu}\partial_{\mu}\nu_l
+u_{l}^{\dagger}i\tilde{\tau}_{\mu}\partial_{\mu}u_{l}+
$$
\begin{equation}
+e_r^{\dagger}i\tau_{\mu}\partial_{\mu}e_r +
d_r^{\dagger}i\tau_{\mu}\partial_{\mu}d_r
+u_r^{\dagger}i\tau_{\mu}\partial_{\mu}u_r + L_{\infty}^{int}(A_{\mu},Z_{\mu}),
\label{15-dop}
\end{equation}
where
$$
L_{\infty}^{int}(A_{\mu},Z_{\mu}) 
=\frac{g}{2\cos \theta_w} \nu_l^{\dagger}\tilde{\tau}_{\mu}Z_{\mu}\nu_l
+\frac{2e}{3}u_{l}^{\dagger}\tilde{\tau}_{\mu}A_{\mu}u_{l} +
$$
$$
+g'\sin \theta_w e_r^{\dagger}\tau_{\mu}Z_{\mu}e_r 
+\frac{g}{\cos \theta_w}\left(\frac{1}{2}-\frac{2}{3}\sin^2\theta_w\right) u_{l}^{\dagger}\tilde{\tau}_{\mu}Z_{\mu}u_{l}
 - g'\cos \theta_w e_r^{\dagger}\tau_{\mu}A_{\mu}e_r-
$$
\begin{equation} 
-\frac{1}{3}g'\cos\theta_w d_r^{\dagger}\tau_{\mu}A_{\mu}d_r + \frac{1}{3}g'\sin\theta_w d_r^{\dagger}\tau_{\mu}Z_{\mu}d_r 
 + \frac{2}{3}g'\cos\theta_w u_r^{\dagger}\tau_{\mu}A_{\mu}u_r
-\frac{2}{3}g'\sin\theta_w u_r^{\dagger}\tau_{\mu}Z_{\mu}u_r.
\label{14}
\end{equation}
We can conclude that
the limit model includes only {\it massless particles}:
photons $A_{\mu}$ and neutral  bosons $Z_{\mu}$, left quarks $u_{l}$ and neutrinos $\nu_l$, right electrons $e_r $ and  quarks $ u_r, d_r $.
 This  phenomenon has simple physical explanation: the  temperature is so high, that particle mass becomes negligible quantity as compared to its kinetic energy.
 The electroweak interactions become long-range because they are mediated by the  massless neutral $Z$-bosons and photons.
Let us note that $W^{\pm}_{\mu}$-boson fields, which  correspond to the translation subgroup of Euclid group $E(2)$, are absent in the limit Lagrangian  $L_{\infty}$ (\ref{15-dop}).

Similar high energies can exist in early Universe after inflation and reheating on the first stages of the Hot Big Bang 
\cite{GR-2011,L-1990} in pre-electroweak epoch. 
%
%
However more interesting is the Universe evolution  and
the limit Lagrangian $L_{\infty}$ can be considered as a good approximation near the  Big Bang  
just as the nonrelativistic   mechanics is a good  approximation of the relativistic one at low velocities.

From the explicit form  of the interaction part
$L_{\infty}^{int}(A_{\mu},Z_{\mu})$
it follows that there are no interactions between particles of different kind, for example neutrinos  interact only with each other by  neutral currents. All other particles are charged and interact with particles of the same sort 
by massless $Z_{\mu}$-bosons and photons. Particles of different kind do not interact.
It looks like some stratification of the Electroweak Model with only one sort of particles in each stratum.

From contraction of the Electroweak Model we can classify events in time as earlier-later, but we can not
determine their absolute time without additional assumptions.
Already at the level of classical gauge fields we can  conclude that the $u$-quark  first restores  its mass in the evolution of the Universe.
Indeed the mass term of $u$-quark in the Lagrangian (\ref{15})
$ L_1=-m_u(u_r^{\dagger}u_{l} + u_{l}^{\dagger}u_r) $
is proportional to the first power $\epsilon  $,
whereas the mass terms of $Z$-boson, electron and $d$-quark are multiplied by the second power of the contraction parameter
\begin{equation}
\epsilon^2\, \left[\frac{1}{2}m_Z^2\left(Z_\mu\right)^2 + m_e(e_r^{\dagger}e_l + e_l^{\dagger} e_r)+
 m_d(d_r^{\dagger}d_{l} + d_{l}^{\dagger}d_r)\right]. 
\label{40}
\end{equation}
At the same time massless Higgs boson $\chi$ and charged $W$-boson are appeared. They  restore their masses after all other particles of the Electroweak Model because their mass terms are multiplied by $\epsilon^4 $.

The electroweak interactions between elementary particles are  restored mainly in the epoch which corresponds to the second order of the contraction parameter. There is one term in Lagrangian (\ref{8})
$L_3=gW_\mu^+W_\mu^-\chi$  proportionate to $\epsilon^3$. The final reconstruction of the electroweak interactions takes place at the last stage ($\approx\epsilon^4 $) together with restoration of mass of all particles.

Two other generations of leptons  and quarks  are developed in a similar way:
for infinite energy there are only massless right $\mu$- and $\tau$-muons, left $\mu$- and $\tau$-neutrinos, as well as massless left and right quarks $c_l, c_r, s_r, t_l, t_r, b_r$. $c$- and $t$-quarks first acquire their mass and after that $\mu$-,  $\tau$-muons, $s$-,  $b$-quarks  become massive.



\section{QCD with contracted gauge group}

Strong interactions of quarks are described by the QCD.
Like the Electroweak Model QCD is a gauge theory based on the local color degrees of freedom \cite{Em-2007}.
The QCD  gauge group is $SU(3)$, acting in three dimensional complex space ${\bf C}_3$ of  color quark states.
 The $SU(3)$ gauge bosons are called gluons. There are eight gluons in total, which
 are the force carrier of the theory between quarks. The QCD Lagrangian is taken in the form
\begin{equation}
{\cal L} =\sum_q \bar{q}^i(i\gamma^\mu)(D_\mu)_{ij}q^j
-\frac{1}{4}\sum_{\alpha=1}^8
F_{\mu\nu}^\alpha F^{\mu\nu\, \alpha},
 \label{q1}
\end{equation}
where $D_\mu q$ are covariant derivatives of quark fields $q=u, d, s, c, b, t $
\begin{equation}
D_\mu q=  
\left(\partial_{\mu}-ig_s\left(\frac{\lambda^\alpha}{2}\right)A^\alpha_{\mu}\right)q,\quad
q=
\left(\begin{array}{c}
 q_1\\
 q_2 \\
 q_3
 \end{array}
 \right) \equiv
\left(\begin{array}{c}
 q_R\\
 q_G \\
 q_B
 \end{array}
 \right)
 \in {\bf C}_3,
 \label{q3}
\end{equation}
$g_S$ is  the strong coupling constant, $t^a=\lambda^a/2$ are generators of $SU(3)$, $\lambda^a $ are Gell-Mann matrices
in the form
$$
\lambda^1=
\left(\begin{array}{ccc}
 \cdot & 1      & \cdot  \\
   1   & \cdot  & \cdot  \\
 \cdot & \cdot  & \cdot  \\
 \end{array}
 \right), \quad
\lambda^2=
\left(\begin{array}{ccc}
 \cdot & -i      & \cdot  \\
   i   & \cdot  & \cdot  \\
 \cdot & \cdot  & \cdot  \\
 \end{array}
 \right), \quad
 \lambda^3=
\left(\begin{array}{ccc}
    1   & \cdot  & \cdot  \\
 \cdot  & -1     & \cdot  \\
 \cdot  & \cdot  & \cdot  \\
 \end{array}
 \right), 
 $$
 $$
\lambda^4=
\left(\begin{array}{ccc}
 \cdot & \cdot  & 1  \\
 \cdot & \cdot  & \cdot  \\
   1   & \cdot  & \cdot  \\
 \end{array}
 \right), \quad
%
\lambda^5=
\left(\begin{array}{ccc}
 \cdot & \cdot  & -i     \\
 \cdot & \cdot  & \cdot  \\
  i    & \cdot  & \cdot  \\
 \end{array}
 \right), \quad
\lambda^6=
\left(\begin{array}{ccc}
 \cdot &  \cdot  & \cdot  \\
   \cdot   & \cdot  & 1   \\
 \cdot & 1  & \cdot  \\
 \end{array}
 \right), 
 $$
  \begin{equation}
 \lambda^7=
\left(\begin{array}{ccc}
    \cdot   & \cdot  & \cdot  \\
 \cdot  & \cdot     & -i  \\
 \cdot  & i  & \cdot  \\
 \end{array}
 \right), \quad
\lambda^8=\frac{1}{\sqrt{3}}
\left(\begin{array}{ccc}
 1 & \cdot  & \cdot  \\
 \cdot & 1  & \cdot  \\
   \cdot   & \cdot  & -2  \\
 \end{array}
 \right),
 \label{q3-1}
\end{equation}
gluon stress tensor
\begin{equation}
F_{\mu\nu}^\alpha=\partial_{\mu} A_\nu^\alpha-\partial_{\nu} A_\mu^\alpha+
g_sf^{\alpha\beta\gamma}A_\mu^\beta A_\nu^\gamma,
 \label{q3-2}
\end{equation}
with the nonzero antisymmetric on all indices structure constant of the gauge group:
$$
f^{123}=1,\quad f^{147}=f^{246}=f^{257}=f^{345}=\frac{1}{2},
$$
\begin{equation}
f^{156}=f^{367}= -\frac{1}{2},\quad
f^{458}=f^{678}=\frac{\sqrt{3}}{2},
 \label{q3-3}
\end{equation}
where $[t^\alpha,t^\beta]=if^{\alpha\beta\gamma}t^\gamma,\; \alpha, \beta, \gamma=1,\ldots,8$.
Mass terms  $-m_q\bar{q}^iq_i$ are not included as far as they are present in the electroweak Lagrangian. 

The choice of Gell-Mann matrices in the form (\ref{q3-1})  fix the basis in $SU(3)$. This enable us to write out  the covariant derivatives (\ref{q3}) in the explicit form
$$
D_\mu={\mathbf I}\partial_\mu -i\frac{g_s}{2}
\left(\begin{array}{cccccccc}
 A_\mu^3+\frac{1}{\sqrt{3}}A_\mu^8 & A_\mu^1-i A_\mu^2 & A_\mu^4-i A_\mu^5\\
 A_\mu^1+iA_\mu^2 & \frac{1}{\sqrt{3}}A_\mu^8- A_\mu^3 & A_\mu^6-i A_\mu^7\\
 A_\mu^4+iA_\mu^5 & A_\mu^6+i A_\mu^7 & -\frac{2}{\sqrt{3}} A_\mu^8 \\
 \end{array}
 \right)=
 $$
 \begin{equation}
={\mathbf I}\partial_\mu
 -i\frac{g_s}{2}
\left(\begin{array}{cccccccc}
 A_\mu^{RR} & A_\mu^{RG} & A_\mu^{RB} \\
 A_\mu^{GR} & A_\mu^{GG} & A_\mu^{GB} \\
 A_\mu^{BR} & A_\mu^{BG} &  A_\mu^{BB} \\
 \end{array}
 \right),
 \label{q4}
\end{equation}
where
$$
A_\mu^{RR}=\frac{1}{\sqrt{3}}A_\mu^8+A_\mu^3, \quad A_\mu^{GG}= \frac{1}{\sqrt{3}}A_\mu^8- A_\mu^3,\quad
A_\mu^{BB}= -\frac{2}{\sqrt{3}} A_\mu^8,
$$
$$
A_\mu^{RR}+ A_\mu^{GG}+ A_\mu^{BB}=0, \quad
A_\mu^{GR}= A_\mu^1+iA_\mu^2 =\bar{A}_\mu^{RG},
$$
\begin{equation}
A_\mu^{BR}= A_\mu^4+iA_\mu^5 =\bar{A}_\mu^{RB}, \quad
A_\mu^{BG}= A_\mu^6+iA_\mu^7 =\bar{A}_\mu^{GB}.
 \label{q4-1}
\end{equation}

 Let us note, that in QCD  the special mechanism of spontaneous symmetry breaking is absent, 
 therefore gluons are massless particles.


The contracted special unitary group $SU(3;\kappa)$ is defined
by the action
$$
q'(\kappa)=\left(\begin{array}{c}
 q'_{1}\\
 \kappa_1 q'_{2} \\
 \kappa_1\kappa_2 q'_{3}
 \end{array}
 \right)=
\left(\begin{array}{ccc}
 u_{11}  &\kappa_1 u_{12} &\kappa_1\kappa_2 u_{13} \\
 \kappa_1 u_{21} & u_{22} & \kappa_2 u_{23} \\
 \kappa_1\kappa_2 u_{31} & \kappa_2 u_{32} & u_{33}
 \end{array}
 \right)
 \left(\begin{array}{c}
 q_{1}\\
 \kappa_1 q_{2} \\
 \kappa_1\kappa_2 q_{3}
 \end{array}
 \right)=
 $$
 \begin{equation}
=U(\kappa )q(\kappa ), \quad  \det U(\kappa )=1, \quad U(\kappa )U^{\dagger}(\kappa )=1
 \label{q7}
\end{equation}
on the complex space ${\bf C}_3(\kappa )$ in such a way that   the hermitian form
\begin{equation}
 q^{\dagger}(\kappa )q(\kappa )= \left|q_1\right|^2+  \kappa_1^2\left(\left|q_2\right|^2+  \kappa_2^2 \left|q_3\right|^2\right)
 \label{q8}
\end{equation}
remains  invariant,
when the contraction parameters  tend to zero:
$\kappa_1, \kappa_2 \rightarrow 0$.
Transition from the classical group $SU(3)$ and space ${\bf C}_3$ to the group $SU(3;\kappa)$ and space ${\bf C}_3(\kappa)$ is given by the substitution
$$
q_1 \rightarrow q_1, \quad
q_2\rightarrow \kappa_1  q_2,\quad q_3\rightarrow \kappa_1\kappa_2  q_3,
$$
\begin{equation}
A_\mu^{GR}\rightarrow \kappa_1 A_\mu^{GR},\quad
A_\mu^{BG}\rightarrow \kappa_2 A_\mu^{BG},\quad
A_\mu^{BR}\rightarrow \kappa_1 \kappa_2 A_\mu^{BR},
 \label{q9}
\end{equation}
and diagonal gauge fields
$A_\mu^{RR}, A_\mu^{GG}, A_\mu^{BB} $
remain unchanged. 

Substituting (\ref{q9}) in (\ref{q1}), we obtain the quark part of Lagrangian in the form
$$
{\cal L}_q(\kappa )= \sum_q \Biggl\{ i\bar{q}_1\gamma^\mu\partial_{\mu} q_1
+\frac{g_s}{2} \left|q_1\right|^2 \gamma^\mu A_\mu^{RR}+
$$
$$
+\kappa_1^2 \biggl[
i\bar{q}_2\gamma^\mu \partial_{\mu} q_2 +
\frac{g_s}{2}\biggl(\left|q_2\right|^2 \gamma^\mu A_\mu^{GG}+
q_1\bar{q}_2\gamma^\mu A_\mu^{GR}+ \bar{q}_1q_2\gamma^\mu \bar{A}_\mu^{GR}\biggr) \biggr] +
$$
$$
+\kappa_1^2 \kappa_2^2 \biggl[
i\bar{q}_3\gamma^\mu \partial_{\mu} q_3 +
\frac{g_s}{2}\biggl(\left|q_3\right|^2 \gamma^\mu A_\mu^{BB}+
q_1\bar{q}_3\gamma^\mu A_\mu^{BR}+ \bar{q}_1q_3\gamma^\mu \bar{A}_\mu^{BR}+
$$
\begin{equation}
+ q_2\bar{q}_3\gamma^\mu A_\mu^{BG}+ \bar{q}_2q_3\gamma^\mu \bar{A}_\mu^{BG}
\biggr) \biggr]  \Biggr\}
 = L_q^{\infty} + \kappa_1^2 L_q^{(2)} + \kappa_1^2\kappa_2^2 L_q^{(4)}.
 \label{q10}
\end{equation}
Let us introduce the notations
 \begin{equation}
\partial A^k\equiv \partial_{\mu} A_\nu^k-\partial_{\nu} A_\mu^k, \quad
[k,m]\equiv A_\mu^k A_\nu^m-A_\mu^m A_\nu^k,
 \label{q11-AA}
\end{equation}
then the gluon tensor  has the following components
$$
F_{\mu\nu}^1=\kappa_1 \biggl\{\partial A^1+
\frac{g_s}{2} \left( 2[2,3]
+\kappa_2^2 \left([4,7] -[5,6] \right) \right) \biggr\},
$$
$$
F_{\mu\nu}^2=\kappa_1 \biggl\{\partial A^2+
\frac{g_s}{2} \left( -2[1,3]
+\kappa_2^2 \left([4,6] +[5,7] \right) \right) \biggr\},
$$
$$
F_{\mu\nu}^3=\partial A^3+
\frac{g_s}{2}\left(\kappa_1^2 2[1,2]
-\kappa_2^2[6,7]
+\kappa_1^2\kappa_2^2[4,5] \right),
$$
$$
F_{\mu\nu}^4=\kappa_1 \kappa_2 \biggl\{\partial A^4 -
\frac{g_s}{2}\left([1,7] +[2,6] +[3,5] -
\sqrt{3}[5,8]\right) \biggr\},
$$
$$
F_{\mu\nu}^5=\kappa_1 \kappa_2 \biggl\{\partial A^5 +
\frac{g_s}{2}\left([1,6] -[2,7] +[3,4] -
\sqrt{3}[4,8]\right) \biggr\},
$$
$$
F_{\mu\nu}^6=  \kappa_2 \biggl\{ \partial A^6 +
\frac{g_s}{2}\left(
\kappa_1^2\left([2,4] -[1,5]\right) +[3,7]+
\sqrt{3}[7,8]\right) \biggr\},
$$
$$
F_{\mu\nu}^7=  \kappa_2 \biggl\{ \partial A^7 +
\frac{g_s}{2}\left(
\kappa_1^2\left([1,4] +[2,5]\right) -[3,6]-
\sqrt{3}[6,8]\right) \biggr\},
$$
\begin{equation}
F_{\mu\nu}^8=\partial A^8
+\frac{g_s\sqrt{3}}{2}\kappa_2^2\left(
\kappa_1^2[4,5]+[6,7]\right).
 \label{q11}
\end{equation}
The gluon part of Lagrangian  is as follows
$$
{\cal L}_{gl}(\kappa)=-\frac{1}{4}F_{\mu\nu}^\alpha F^{\mu\nu\, \alpha}=
$$
$$
=-\frac{1}{4}\Biggl\{ H_3^2+H_8^2 + \kappa_1^2\left(F_1^2+ F_2^2 + 2H_3F_3 \right) +
$$
$$
+\kappa_2^2 \left(G_6^2+ G_7^2+ 2H_3G_3 -2\sqrt{3}H_8G_3 \right) + \kappa_1^4F_3^2 +\kappa_2^4 4G_3^2 +
$$
$$
+\kappa_1^2\kappa_2^2\left[P_4^2+P_5^2 +2\left( F_1G_1 +F_2G_2 +F_3G_3 +F_6G_6 +F_7G_7 +\right. \right.
$$
$$
\left. \left.
+ \sqrt{3}H_8P_3\right) \right]+ \kappa_1^2\kappa_2^4\left(G_1^2 +G_2^2  -4G_3P_3  \right) +
$$
 \begin{equation}
+ \kappa_1^4\kappa_2^2\left(F_6^2 +F_7^2 +2F_3P_3 \right) +
\kappa_1^4\kappa_2^4 4P_3^2\Biggr\}.
 \label{q11-A}
\end{equation}
where
$$
F_1=\partial A^1 +g_s[2,3], \quad
F_2=\partial A^2 -g_s[1,3],
$$
$$
G_1=\frac{g_s}{2}\left( [4,7]-[5,6]  \right), \quad
G_2=\frac{g_s}{2}\left( [4,6]+[5,7]  \right),
$$
$$
H_3= \partial A^3, \quad
F_3= g_s [1,2], \quad
G_3= -\frac{g_s}{2} [6,7], \quad
P_3= \frac{g_s}{2} [4,5],
$$
$$
P_4=\partial A^4-
\frac{g_s}{2}\left([1,7]+[2,6]+[3,5]-\sqrt{3}[5,8] \right),
$$
$$
P_5=\partial A^5+
\frac{g_s}{2}\left([1,6]-[2,7]+[3,4]-\sqrt{3}[4,8] \right),
$$
$$
G_6=\partial A^6+
\frac{g_s}{2}\left([3,7]+\sqrt{3}[7,8] \right), \quad
F_6=\frac{g_s}{2}\left([2,4]-[1,5] \right),
$$
$$
G_7=\partial A^7-
\frac{g_s}{2}\left([3,6]+\sqrt{3}[6,8] \right), \quad
F_7=\frac{g_s}{2}\left([1,4]+[2,5] \right),
$$
 \begin{equation}
 H_8= \partial A^8.
\label{q11-AB}
\end{equation}

In the framework of Cayley-Klein scheme  \cite{Gr-12}
the gauge group $SU(3;\kappa)$ has two one-parameter contractions $\kappa_1 \rightarrow 0, \kappa_2=1 $ and
$ \kappa_2 \rightarrow 0, \kappa_1=1$, as well as one two-parameter contraction
$\kappa_1, \kappa_2 \rightarrow 0$.
We consider the following contraction: $\kappa_1=\kappa_2=\kappa=\epsilon \rightarrow 0$, which corresponds to the infinite temperature limit  of QCD.
The quark part of Lagrangian (\ref{q10}) is represented as a sum of terms proportional to zero, the second and forth powers of contraction parameter $\epsilon$
\begin{equation}
{\cal L}_q(\epsilon )= L_q^{\infty} + \epsilon^2 L_q^{(2)} + \epsilon^4 L_q^{(4)}
 \label{q11-C}
\end{equation}
and gluon part   is represented as a sum 
\begin{equation}
{\cal L}_{gl}(\epsilon)=
L_{gl}^{\infty} + \epsilon^2 L_{gl}^{(2)} + \epsilon^4 L_{gl}^{(4)}+ \epsilon^6 L_{gl}^{(6)}+ \epsilon^8 L_{gl}^{(8)}.
 \label{q11-D}
\end{equation}

In the infinite temperature limit
$\kappa=\epsilon \rightarrow 0$
the most parts of gluon tensor components are equal to zero
and the expressions for two nonzero components are simplified
\begin{equation} 
F_{\mu\nu}^3=\partial_{\mu} A_\nu^3-\partial_{\nu} A_\mu^3=\frac{1}{2}\left(F_{\mu\nu}^{RR}-F_{\mu\nu}^{GG}\right),\quad
F_{\mu\nu}^8=\partial_{\mu} A_\nu^8-\partial_{\nu} A_\mu^8=\frac{\sqrt{3}}{2}\left(F_{\mu\nu}^{RR}+F_{\mu\nu}^{GG}\right),
 \label{q12}
\end{equation}
so we can  write out the QCD Lagrangian ${\cal L}_{\infty}$ in this limit explicitly
$$
{\cal L}_{\infty}=L_q^{\infty}+L_{gl}^{\infty}=
$$
 \begin{equation}
=\sum_q \biggl\{ i\bar{q}_R\gamma^\mu\partial_{\mu} q_R
+\frac{g_s}{2} \left|q_R\right|^2 \gamma^\mu A_\mu^{RR} \biggr\} 
-\frac{1}{4}\left(F_{\mu\nu}^{RR}\right)^2
-\frac{1}{4}\left(F_{\mu\nu}^{GG}\right)^2
-\frac{1}{4}F_{\mu\nu}^{RR}F_{\mu\nu}^{GG}.
 \label{q13}
\end{equation}
  From ${\cal L}_{\infty}$ we conclude that only the dynamic terms for the first color component of massless quarks survive under infinite temperature, which means that the quarks are monochromatic, and  the terms also survive, which describe the interactions of these components with $R$-gluons.
Besides $R$-gluons there are also $G$-gluons, which do not interact with the quarks.

Similarly to the Electroweak Model starting with ${\cal L}_q(\epsilon ) $ (\ref{q11-C}) and ${\cal L}_{gl}(\epsilon ) $ (\ref{q11-D}) one can construct a number of intermediate models for QCD, which describe the gradual restoration of color degrees of freedom for the quarks and the gluon interactions in the Universe evolution.

It follows from Lagrangian
${\cal L}(\epsilon )={\cal L}_q(\epsilon ) +
{\cal L}_{gl}(\epsilon ) $,
that the total reconstruction of the quark color degrees of freedom will take place after the restoration of all quark mass $(\approx \epsilon^2)$ at the same time with the reestablishment of all electroweak interactions $(\approx \epsilon^4)$. Complete color interactions start  to work later because some of  them are proportionate to the eighth  power $\epsilon^8 $.

\section{ Estimation  of boundary values} 

As it was mentioned the contraction of gauge group of  QCD gives an opportunity to order in time different stages of its development, but does not make it possible to  bear  their absolute  date.
Let us try to estimate this date with the help of additional assumptions.  The equality of the contraction parameters for QCD and the EWM 
is one of these assumptions.

Then we use the fact  that the electroweak epoch starts at the temperature  $T_4=100\, GeV\;$ $(1\, GeV=10^{13} K)$ and the QCD epoch begins at $T_8=0,2\, GeV$. In other words we assume that complete reconstruction of the EWM, 
whose Lagrangian has minimal terms proportionate to $\epsilon^4$, and QCD, whose Lagrangian has minimal terms proportionate to $\epsilon^8$, take place at these temperatures.
Let us denote by $\Delta$  cutoff  level for 
$\epsilon^k, \; k=1,2,4,6,8$, i.e. for $\epsilon^k < \Delta$ all the terms  proportionate to $\epsilon^k$  are negligible quantities in Lagrangian.
At last we suppose that the contraction parameter  inversely depends on temperature 
\begin{equation}
\epsilon(T)=\frac{A}{T},
 \label{q14}
\end{equation}
where $A$ is constant.

As far as the minimal terms in the QCD Lagrangian are proportional to  $\epsilon^8$ and QCD is completely  reconstructed at $T_8=0,2\, GeV$, we have the equation
$\epsilon^8(T_8)=A^8T_8^{-8}= \Delta $
and obtain  $A=T_8\Delta^{1/8}=0,2\Delta^{1/8}\, GeV$.
The minimal terms in 
the EWM Lagrangian are proportional to  $\epsilon^4$ and it is  reconstructed at $T_4=100\, GeV$, so we have
$\epsilon^4(T_4)=A^4T_4^{-4}=\Delta $, i.e.
$T_4=A\Delta^{-1/4}=T_8\Delta^{1/8}\Delta^{-1/4}=T_8\Delta^{-1/8}$
and we obtain the  cutoff  level 
$\Delta = (T_8T_4^{-1})^8=(0,2\cdot10^{-2})^8\approx 10^{-22}$,
which is consistent with the typical energies of the Standard Model.
From the equation
$\epsilon^k(T_k)=A^kT_k^{-k}=\Delta $
we obtain
\begin{equation}
T_k=\frac{A}{\Delta^{1/k}}=\frac{T_8\Delta^{1/8}}{\Delta^{1/k}}= T_8\Delta^{\frac{k-8}{8k}}\approx 10^{\frac{88-15k}{4k}}\, GeV.
 \label{q15}
\end{equation}
Simple calculations give the following estimations for the boundary values of the  temperature in the early Universe  ($GeV $):
$T_1=10^{18},\; T_2=10^7,\; T_3=10^3,\; T_4=10^2,\; T_6=1,\; T_8=2\cdot10^{-1}$.
The obtained estimation for "infinity" temperature 
$T_1\approx 10^{18}\, GeV $
is comparable with Planck energy
$\approx 10^{19}\,  GeV $, where the gravitation effects are important.
So the developed evolution of the elementary particles does not exceed the range of the problems described by electroweak and strong interactions.

It should be noted that for the power function class
 $ \epsilon(T)=BT^{-p}  $
the estimations for temperature  boundary values are very weakly dependent on power $p$. So for $p=10$ we obtain practically the same $T_k$ as for the simplest function (\ref{q14}) with $p=1$.

\section*{Conclusion}
\addcontentsline{toc}{section}{Conclusion}

We have investigated   the high-temperature  limit of the SM
which was obtained from the  first principles of the gauge theory as contraction of its gauge group.
It was shown that the mathematical contraction parameter is proportional   to  the temperature  and its zero limit corresponds to the infinite temperature limit of the  Model.
The SM passes in this limit through several stages, which are distinguished by the powers of contraction parameter,
what gives the opportunity to classify them in time as earlier-later. To determine the absolute date of these  stages the additional assumptions was used, namely: the inversely dependence $\epsilon $ on the temperature  (\ref{q14}) and  the cutoff  level $\Delta $ for $\epsilon^k $. Unknown parameters are determined with the help of the QCD and EWM typical energies.

The exact expressions for the respective Lagrangians for any stage in the SM evolution are presented. 
On the base of decompositions (\ref{15}), (\ref{q11-C}), (\ref{q11-D})  the intermediate models ${\cal L}_{k} $ for any temperature  scale are constructed.
It gives an opportunity to draw a conclusions on interactions and properties of elementary particles in each of considered epoch.
The presence of the several intermediate models in the interval  from Plank energy $10^{19}\, GeV $ up to the EWM typical energy $10^{2}\, GeV $ instead of only one model  automatically takes away    so-called  hierarchy problem of the SM \cite{Em-2007}.

At the infinite temperature  limit
($T> 10^{18}\,  GeV$)
 all particles including vector bosons lose their masses and
electroweak interactions become long-range.
Monochromatic massless quarks exchange by only one sort of $R$-gluons. Besides $R$-gluons there are also $G$-gluons, which do not interact with quarks.
It follows from the explicit form of Lagrangians
$L_{\infty}^{int}(A_{\mu},Z_{\mu})$ (\ref{14}) and ${\cal L}_{\infty} $ (\ref{q13})
that only the particles of the same sort  interact with each other. Particles of different sorts do not interact.
It looks like some stratification of leptons and quark-gluon plasma with only one sort of particles in each stratum.

 At the level of classical gauge fields it is already possible to give some conclusions on the appearance   of elementary particles mass  on the different stages of the Universe evolution.
In particular we can  conclude that half of quarks
$(\approx\epsilon,\; 10^{18}\,  GeV > T > 10^7\,  GeV) $
 first restore  they mass.
Then $Z$-bosons, electrons and other quarks become massive
$(\approx \epsilon^2,\; 10^{7}\, GeV > T > 10^3\, GeV) $.
Finally  Higgs boson $\chi$ and charged $W^{\pm}$-bosons restore their masses because their mass terms are multiplied by $\epsilon^4\;$ ($ T < 10^2 \, GeV$).

In a similar way it is possible to describe the evolution of particle interactions.
 Self-action of Higss boson appears with its mass restoration.
At the same epoch start interactions of four  $W^{\pm}$-bosons, as well as of two Higgs  and two $W$-bosons (\ref{8-1}).
The only one term in Lagrangian, which is proportional to the third power of $\epsilon$, describes interaction of Higss boson with charged $W^{\pm}$-bosons
$( T < 10^3\, GeV) $.
The rest of the electroweak particle interactions are appeared in the second order of the contraction parameter
$(10^{7}\, GeV > T > 10^3\, GeV) $.
Some part of color interactions between quarks in Lagrangian
(\ref{q10}) is proportional to $\epsilon^2\;$
($T <10^{7}\, GeV$)
 and the rest part is proportional to
$\epsilon^4\;$
($T <10^{2}\, GeV$).
Therefore the complete restoration of quark color degrees of freedom takes place after the appearance of quark masses
 ($\approx \epsilon^2,\; T <10^{7}\, GeV$)
 (\ref{40})
together with the restoration of all electroweak interactions
($\approx \epsilon^4,\; T <10^{2}\, GeV$).
Comple\-te color interactions start later because they are proportional to
 $\epsilon^8\;$ ($T <10^{-1}\, GeV$).

The evolution of elementary particles and their interactions in the early Universe obtained with the help of contractions of gauge groups of the SM does not contradict  the canonical one  \cite{GR-2011}, according to which the QCD phase transitions take place later then the electroweak phase transitions. The developed evolution of the SM present the basis for a more detailed analysis of different phases in the formation of leptons and quark-gluon plasma.

The author is thankful to V.~V.~Kuratov  and V.~I.~Kostyakov for helpful discussions.
The study is supported by Program of UrD RAS project N~15-16-1-3.



\end{document}